\documentclass[12pt]{spieman}

\usepackage{silence}
\usepackage{graphicx} 
\usepackage{xcolor} 
\usepackage{csvsimple}
\usepackage{siunitx}
\usepackage{placeins}
\usepackage{subcaption}
\usepackage{comment}
\usepackage[left]{lineno}

\expandafter\let\csname equation*\endcsname\relax
\expandafter\let\csname endequation*\endcsname\relax
\usepackage{amsmath}

\newcommand{\figref}[1]{\figurename~\ref{#1}}
\newcommand{\secref}[1]{Section~\ref{#1}}
\newcommand{\tabref}[1]{\tablename~\ref{#1}}
\NewDocumentCommand{\rot}{O{90} O{0.25em} m}{\makebox[#2][l]{\rotatebox{#1}{#3}}}%

\DeclareMathOperator{\NRMSE}{NRMSE}
\DeclareMathOperator{\argmin}{argmin}
\DeclareMathOperator{\MedAPE}{MedAPE}
\DeclareMathOperator{\median}{median}

\graphicspath{ {./Figures/} }
\ActivateWarningFilters[csv]

\DeactivateWarningFilters[csv]

\usepackage{hyperref}

\usepackage{marginnote}

\title{Synthetic white balancing for intra-operative hyperspectral imaging}
\author[a, *]{Anisha Bahl}
\author[a, b]{Conor C Horgan}
\author[b]{Mirek Janatka} 
\author[a, c]{Oscar J MacCormac} 
\author[b]{Philip Noonan} 
\author[a]{Yijing Xie} 
\author[a, e]{Jianrong Qiu}
\author[d]{Nicola Cavalcanti} 
\author[d]{Philipp F\"urnstahl} 
\author[b]{Michael Ebner}
\author[e]{Mads S Bergholt} 
\author[a, b, c]{Jonathan Shapey}
\author[a, b]{Tom Vercauteren}

\affil[a]{School of Biomedical Engineering \& Imaging Sciences, King's College London, 1 Lambeth Palace Road, London, United Kingdom}
\affil[b]{Hypervision Surgical Ltd, 85 Great Portland Street, London, United Kingdom}
\affil[c]{King's College Hospital, Denmark Hill, London, United Kingdom}
\affil[d]{Balgrist University Hospital, Forchstrasse 340, 8008 Zurich, Switzerland}
\affil[e]{Centre for Craniofacial and Regenerative Biology, King's College London, London, United Kingdom}

\begin{document} 
\maketitle
\noindent \footnotesize\textbf{*}Corresponding author,  \linkable{anisha.bahl@kcl.ac.uk} 

\begin{abstract}

\noindent \textbf{Purpose} 

\noindent Hyperspectral imaging shows promise for surgical applications to non-invasively provide spatially-resolved, spectral information.
For calibration purposes,
a white reference image of a highly-reflective Lambertian surface should be obtained under the same imaging conditions.
Standard white references are not sterilizable, and so are unsuitable for surgical environments.
We demonstrate the necessity for in situ white references and address this by proposing a novel, sterile, synthetic reference construction algorithm.

\noindent \textbf{Approach}

\noindent The use of references obtained at different distances and lighting conditions to the subject were examined.
Spectral
and color reconstructions
were compared
with standard measurements qualitatively and quantitatively, using $\Delta E$ and normalised RMSE respectively. 
The algorithm forms a composite image from a video of a standard sterile ruler, whose imperfect reflectivity is compensated for. The reference is modelled as the product of independent spatial and spectral components, and a scalar factor accounting for gain, exposure, and light intensity. 
Evaluation of synthetic references against ideal but non-sterile references is performed using the same metrics alongside pixel-by-pixel errors. 
Finally, intraoperative integration is assessed though cadaveric experiments.

\noindent \textbf{Results}

\noindent Improper white balancing leads to increases in all quantitative and qualitative errors. 
Synthetic references achieve median pixel-by-pixel errors lower than 6.5\%  and produce similar reconstructions and errors to an ideal reference.
The algorithm integrated well into surgical workflow, achieving median pixel-by-pixel errors of 4.77\%, while maintaining good spectral and color reconstruction.

\noindent \textbf{Conclusions} 

\noindent We demonstrate the importance of in situ white referencing and present a novel synthetic referencing algorithm. This algorithm is suitable for surgery whilst maintaining the quality of classical data reconstruction.
\end{abstract}

\section{Introduction}
\label{intro}
Hyperspectral imaging (HSI) has shown potential in pre-clinical and clinical studies to non-invasively provide information for disease diagnosis and surgical guidance \cite{Lu2014,Giannoni2018,Calin2014,Shapey2019}. HSI provides multi-channel spectral imaging data
across a wide field of view, where each channel represents a narrow spectral measurement centred around a given wavelength. This technique can also be referred to as multispectral imaging when there is a low number of bands, however for simplicity we will refer to this as HSI in all cases.
The diffusely reflected light is measured across a range of wavelengths for each pixel of an image. The diffuse reflection is determined by the scattering properties and absorbing or emitting species in the tissues being examined~\cite{Jacques2013}. This provides the opportunity for non-invasive, quantitative analysis of these tissues to investigate physiologically relevant parameters, such as tissue oxygen saturation \cite{Clancy2020}.

There are three broad categories of HSI cameras: spatial scanning, spectral scanning, and snapshot acquisition. Spatial scanning is typically implemented with a linescan camera which acquires data across all wavelengths simultaneously but scans through each line of pixels sequentially. In contrast, spectral scanning cameras collect all pixels at a given single wavelength band simultaneously using a band pass filter but scan through each wavelength bands sequentially. These methods both measure a complete hypercube directly. This highly resolved spectral data has been used for the majority of medical applications published to date \cite{Kulcke2018a,Giannoni2021,Shapey2019,Yoon2021}.
These HSI approaches are not able to provide real-time data as their respective scanning mechanisms result in long acquisition times that are prone to motion artefacts. In contrast, snapshot mosaic cameras utilize sensors where each pixel has a dedicated band pass filter. This provides information on one band per pixel but captures all pixels in a single shot~\cite{Geelen2014}. This allows highly time resolved data to be obtained, at the cost of lower spatial and spectral resolution. To obtain a full hypercube from a snapshot mosaic camera, the data must be demosaiced with the remaining information inferred using classical or learning-based interpolation \cite{Li2021a}, followed by spectral cross-talk correction due to the parasitic neighboring bands effects \cite{Pichette2017}. The increased temporal resolution given by effective snapshot mosaic image demosaicing provides the opportunity to aid surgical guidance \cite{Ayala2021, Ebner2021}.

To enable physiological parameter extraction from any of these imaging methods, accurate spectra must be extracted from hyperspectral images. This requires white balancing as an initial step. White balancing uses a white and dark reference to account for lighting conditions, vignetting, and optical transmission through the set-up.
The white reference image is typically taken using a well characterized, uniform, highly reflective, Lambertian surface \cite{Lu2014}.
However existing established references, such as 95\% reflective Spectralon tiles, cannot be sterilized and are very sensitive, making them challenging to integrate in a surgical environment.
This necessitates that the white reference is acquired outside 
of the surgical field, resulting in different imaging conditions relative to the subject.
Ideally, a new white reference should be acquired whenever the imaging conditions are altered which, at best, disrupts the clinical workflow and, at worst, is impossible. This forces the settings to be kept constant throughout the procedure, thereby limiting the setup's use, or risks losing the ability to acquire quantitative spectral data. 

To alleviate this issue, \citenum{Khan2017} proposed adaptations of several white balancing models initially developed for standard color photography to determine the light source spectrum from a hyperspectral image.
These models assume a range of colors within the spectral scene to ensure that at least one pixel per band will display the maximum value, or a variation of the grey-world algorithm which assumes the mean of the image will be grey.
These assumptions do not hold well for surgical scenes which often have largely homogenous color schemes.

Spectral and spatial vignetting corrections can also be performed independently. \citenum{Ayala2020} proposed to use
specular highlight analysis for spectral correction.
This requires additional low exposure images to be obtained to prevent saturation in the specular reflection regions.
This allows illuminant spectra to be determined intraoperatively without the need for a reference, however it is unable to directly account for vignetting and can only be used to provide relative data as specular reflections cannot quantitatively be related by Lambertian reflections.
These relative data can be used to extract ratiometric clinical parameters such as oxygen saturation~\cite{Giannoni2018,MacKenzie2018}.
However, this may not allow quantitative measurement of all parameters such as chromophore concentrations, including haemoglobin which is therefore often provided on a relative scale~\cite{Kulcke2018a}.

Vignetting correction can be performed following spectral correction. For hyperspectral images, this assumes a constant optical configuration and requires significant characterisation before use \cite{Jiang2019}. In surgical settings, there are often adjustable focal lengths changing the optical configuration meaning this is challenging to apply. Single image vignetting correction models are available for RGB imaging \cite{Cho2014}, however these also assume natural images with a variety of colors which is not an appropriate assumption for surgical imaging.

Addressing the need for an intraoperative model to account for vignetting and illuminant spectrum whilst allowing quantitative spectral extraction is the focus of this paper, as a theatre-ready solution to this problem has yet to be adopted. 

In this paper, our contributions are threefold: (i) a novel synthetic reference generation algorithm using a widely available standard reference, (ii) quantification of the impact of improper white balancing, and (iii) demonstration of the suitability of these synthetic references in both a controlled and surgical environment. To the best of our knowledge the novelty in the synthetic reference algorithm includes the hyperspectral composite raw image construction approach, the reflectivity correction approach, and the novel formalisation of a separable white reference model. 

This study aims to use snapshot HSI to reproduce known spectra quantitatively using only reference objects widely available in the operating room.
\secref{method} describes the postprocessing steps required for snapshot hyperspectral imaging, and details a novel synthetic reference generation algorithm and the error quantification methods used to evaluate this. \secref{results} quantifies the impact of improper white balancing, followed by the comparison of synthetic references generated using this algorithm to measured non-sterile references in the same conditions alongside their associated reconstructed spectra. Finally, an example of an intra-operative image taken of a human cadaveric spine balanced with both a non-sterile measured reference and a synthetic reference is shown. These results demonstrate a novel algorithm to allow white balancing in a fully sterile environment allowing the recovery of accurate quantitative spectra. 
%

\section{Materials and methods}
\label{method}

%
\subsection{Hyperspectral imaging setup}
\label{materials}
\begin{figure}[htb]
        \centering
        \includegraphics[width=0.9\textwidth]{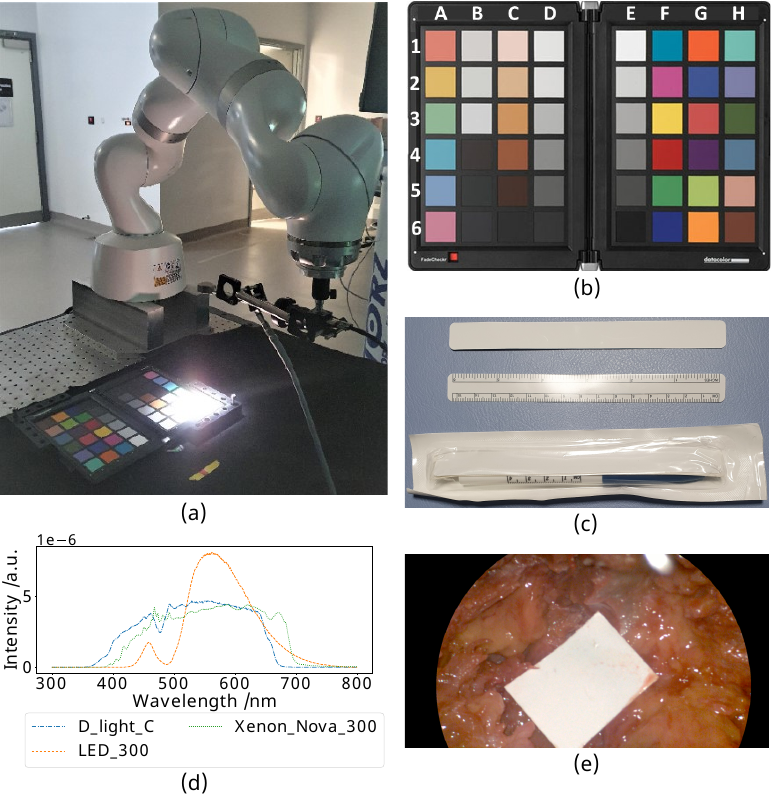}
        \hspace{2mm}
	\caption{(a) Hyperspectral imaging setup mounted to Kuka robot, (b) the colored checkerboard used to provide a variety of well characterized spectra, (c) the sterile ruler used to construct synthetic references, (d) the spectra of the Karl Storz D light C, Karl Storz power LED 300, and Karl Storz Xenon Nova 300 light sources, alongside (e) an sRGB reconstruction of an image of a snipped section of a ruler against tissue taken with the handheld hyperspectral imaging setup whose field of view is cropped due to the scope.}
	\label{fig:setup}
\end{figure}

A snapshot hyperspectral imaging camera with a 4x4 mosaic pattern across the sensor is used in this work.
More specificially, we utilize a
4x4 16 band visible range snapshot mosaic camera (Ximea utilising the IMEC CMV2K-SSM4X4-VIS sensor, Germany) alongside an f=35mm coupler (Karl Storz, NDTec, VCam HD-F-35 - Camera Lens Adapter, Germany) and a $90^\circ$ exoscope (Karl Storz Endoscopy, VITOM Telescope 0° w Integ. Illuminator, UK), as detailed in \citenum{Ebner2021}. An exoscope is chosen to create a standalone imaging system and ease translation to first-in-patient clinical use by minimising disruption to the primary patient care workflow. A disposable sterile ruler commonly found in operating theatres is used as the reference for constructing synthetic references shown in \figref{fig:setup}c.

To capture data in a controlled laboratory environment, allowing quantitative investigation of white referencing significance, we augment our surgical imaging set-up with a robotic mount. The exoscope rotation is fixed and either a Xenon light source (Karl Storz Endoscopy, Cold Light Fountain D light C, UK) or an LED light source (Karl Storz Endoscopy, Power LED 300, UK) is used, whose area normalized emission spectra can be seen in \figref{fig:setup}d. The system is mounted to a seven degree of freedom robot (LBR, KUKA med7 R800, Germany) to enable multiple images to be taken with a constant reproducible camera position relative to the subject.
This is shown in \figref{fig:setup}a. A colored checkerboard target (datacolor, Spyder Checkr, UK), where each tile has a well characterized measured spectrum, is imaged at each distance and shown in \figref{fig:setup}b. 

For seamless integration to the intraoperative environment, the camera is used in a handheld configuration with a Xenon light source (Karl Storz Endoscopy, Xenon Nova 300, UK) whose area normalized emission spectrum can also be seen in \figref{fig:setup}d. A standard Red, Green, Blue (sRGB) reconstruction of an image of a snipped section of this ruler on tissue using this handheld system is shown in \figref{fig:setup}e.

\subsection{Estimating reflectance from hyperspectral data}
\label{methodnecessity}
As detailed by \citenum{Pichette2017}, estimating reflectance from hyperspectral mosaic data necessitates a computational pipeline encompassing white balancing, demosaicing and spectral cross-talk correction.
Our white balancing approach is detailed in the subsequent sections.
For demosaicing purposes, in this work, we rely on bilinear interpolation to maintain the spatial resolution of the raw mosaic image but more advanced approaches such as the learning-based one of \citenum{Li2021a} could also be employed with the work presented here.
The demosaiced hypercube is then corrected pixel by pixel to account for parasitic spectral cross-talk effects and secondary peaks in the band responses within the sensor to return the final hypercube\cite{Pichette2017}.

\subsubsection{White balancing}
\label{processingwhite}
Hyperspectral images must be white balanced before further post-processing to account for the lighting conditions (including light source and illumination distribution), optical transmission through the system, exposure of the camera, quantum efficiency of the sensors, and inherent sensor noise.
Any bias in sensor noise is accounted for by using a dark reference, obtained with the lens cap covering the sensor.
The dark image is then subtracted from the image.
The remainder of the factors are accounted for by using a white reference image, traditionally obtained using a uniformly highly reflective Lambertian surface.
Standard practice is to use a Spectralon 95\% reflective tile, to provide the maximum possible intensity values.
This white image is used to scale the image intensities between 0 and 1.
Given the demosaiced data, this is combined to form a full white balancing step:
\begin{linenomath*}
\begin{equation}
	R(i,j,n) = \frac{\rho_n\left(I(i,j,n) - \frac{\tau_I}{\tau_D}D(i,j,n)\right)}{\frac{\tau_I}{\tau_W}W(i,j,n) - \frac{\tau_I}{\tau_D}D(i,j,n)}
\label{eq:WBfull}
\end{equation}
\end{linenomath*}
where $i$ and $j$ are the spatial co-ordinates and $n$ is the spectral co-ordinate in the uncorrected image $I$, white reference $W$, dark reference $D$, and white-balanced reflectance estimate $R$ hypercubes respectively.
$\tau_I$, $\tau_W$, and $\tau_D$ describe the exposure time used for acquisition of each of the above, and $\rho_n$ is the bandwise reflectivity of the white reference.

To optimize computational performance, white balancing can be performed on the raw mosaic images before demosaicing. When using established white reference materials, the reflectivity $\rho_n$ of each band can be assumed to be a constant $\rho$. These conditions simplify \eqref{eq:WBfull} to:
\begin{linenomath*}
\begin{equation}
    R^m(x,y) = \frac{\rho\left(I^m(x,y) - \frac{\tau_I}{\tau_D}D^m(x,y)\right)}{\frac{\tau_I}{\tau_W}W^m(x,y) - \frac{\tau_I}{\tau_D}D^m(x,y)}
\label{eq:WB}
\end{equation}
\end{linenomath*}
where the superscript $m$ denotes mosaic images with the spatio-spectral coordinates $x$ and $y$.
In \eqref{eq:WB}, spectral bands are implicitly indexed by the spatial coordinates $i$ and $j$ as each pixel in a snapshot mosaic sensor only captures a single spectral band.

In the remainder of this section, we focus on the denominator in \eqref{eq:WBfull} or \eqref{eq:WB}.
For brevity, empirically obtained images or videos will be assumed to be exposure corrected and dark corrected before use, whereas reflectivity correction will be discussed in detail in \secref{algorithmreflectivity}.
We also assume that demosaicing and spectral crosstalk correction as per \citenum{Pichette2017} is applied throughout.

%
\subsection{Quantifying the effect of improper white balancing}
\label{methodmotivation}
As current white references 
can not be used within
a sterile surgical field, changes in distance and lighting conditions are inevitable between the acquisition of patient and calibration images. 
In this section we focus on evaluating the effects of such changes, further considering that these are typically associated with changes in optical focus and low-level camera controls.

An image of each tile of a well-characterized checkerboard, as detailed in \secref{materials}, is acquired under a set of imaging configurations (distance, light source, etc.) and processed to form a hypercube for each tile.
Each tile is annotated with five circles of radius 30 pixels, to ensure the annotations can be performed consistently at different distances where the tile occupies different proportions of the content area. This allows the mean plus or minus the standard deviation of their spectra to be visualized against the spectrometer measurements of that tile to provide a qualitative measure of fit.
This calculation is performed for both the quantitative spectra and for mean normalized relative spectra, as relative spectra can be used to compute some physiologically relevant ratiometric parameters as described in \secref{intro}. 

These datasets were obtained at 7 distances between 20 and 35 cm to cover the comfortable operating range of the exoscope-based set-up in theatre. This was performed using two light sources, Karl Storz D light C and Karl Storz LED. This allows the comparison of a variety of distances and lighting conditions to mimic the differences encountered in sub-optimal clinical white reference conditions to the imaging conditions.

\paragraph{Spectral $\NRMSE$}
For quantitative comparison, the reconstructed spectrum $s_n$ at any given band $n$ is evaluated against the intensity at the same wavelength in the reference spectrum $r_n$ using the normalized root mean squared error ($\NRMSE$) calculated across all $N$ bands:
\begin{linenomath*}
\begin{equation}
    \NRMSE = \frac{\sqrt{\frac{1}{N}\sum_n^N\left(s_n - r_n\right)^2}}{\sqrt{\frac{1}{N}\sum_n^N r^2_n}}
\label{eq:normRMSE}
\end{equation}
\end{linenomath*}

\paragraph{Tristimulus color reconstruction and perceptual color differences}
Additionally, each tile image is converted to sRGB and the centre of each composited to construct an sRGB checkerboard image for a qualitative comparison to the original checkerboard in \figref{fig:setup}b. The same region of each tile can also be converted to CIELAB for pixel-by-pixel comparison to the CIELAB values provided by the manufacturer (\url{https://spyderx.datacolor.com/downloads/SpyderCheckr_Color_Data_V2.pdf}).
To measure the perceptual discrepancy in tristimulus color reconstruction, a $\Delta E$ calculation
is used as is defined in \citenum{Sharma2005}
(using pyciede v0.0.21 \newline \href{https://pypi.org/project/pyciede2000/}{ \texttt{pyciede2000.ciede2000}} function.) A $\Delta E$ value of less than 1 indicates an imperceptible difference in color, while values of up to 6 are considered acceptable for commercial reproduction using printing presses. 

To compute the RGB and CIELAB images, the spectra are first converted to CIEXYZ \cite{Smith1931} using a camera-specific $3 \times N$ matrix.
The CIEXYZ values can be converted to linear sRGB using a literature matrix corresponding to a standard D65 light source
followed by conversion to sRGB using gamma correction to account for standard monitors
\cite{Magnusson2020,Reinhard}
%
%
%
%
%
%
%
%
%
%
%
%
%
%
%

%
%
%
%
%
%
%
%
%
%
%
%
%
%
%
%
%

\subsection{White reference model}
\label{methodmodel}
A measured white reference image $W$ can be demosaiced to a hypercube where there is a full spectrum per spatial pixel $W(i,j,n)$. We propose to factorize the white reference as a product of spatial-only vignetting $V(i,j)$, spectral sensitivities of each band $S(n)$, and a scalar factor $M$ to account for the overall light intensity.
Additionally, noise $N(i,j,n)$ should be accounted for in both the spatial and spectral domains:
\begin{linenomath*}
\begin{equation}
	W(i,j,n) = MS(n)V(i,j) + N(i,j,n)
\label{eq:modelling white references}
\end{equation}
\end{linenomath*}

The assumption of separability is intuitive from the discussion of vignetting and color constancy being addressed separately in literature\cite{Cho2014,Ayala2020,Jiang2019,Yu2004a,Foster2011a}.
By convention, to achieve a well-posed decomposition, we  constrain  it such that $\max_{i,j}V(i,j) = 1$ and $\frac{1}{N} \sum_{n}S(n) = 1$. 
This ensures that: 1) spatial vignetting only accounts for variations in light intensity and optical transmission across the image; 2) spectral sensitivities only account for relative efficiency of the different bands on the sensor and the spectral shape of the light source; and 3) the scalar factor only accounts for electronic gain, exposure, and overall light intensity settings.
For simplicity, this model neglects chromatic aberrations in the optics, and assumes all pixels corresponding to a given band behave the same. 

\subsection{Synthetic white reference algorithm}
\label{methodalgorithm}
Since a sterile standard reference equivalent to a uniformly reflective Lambertian surface is not commercially available, a standard sterile ruler shown in \figref{fig:setup}c is used. This can be placed in the sterile surgical field on the subject to ensure the imaging conditions are identical. The reverse of the ruler has a matt white appearance without markings.
This ruler does not cover the full field of view and so images must be composited to estimate the vignetting over the entire field of view.
To ensure full coverage, a short video of the sterile ruler crossing the field of view is captured to create a synthetic white reference. This allows the readily available, inexpensive reference to be used in a wide variety of operative cavity dimensions.

As compositing introduces imperfections, we exploit constraints in our white reference model to fit a more accurate estimation.
The ruler is also not uniformly reflective across the wavelengths of interest and so this must be accounted for.

\begin{figure}[htb]
	\centering
        \includegraphics[width=\textwidth]{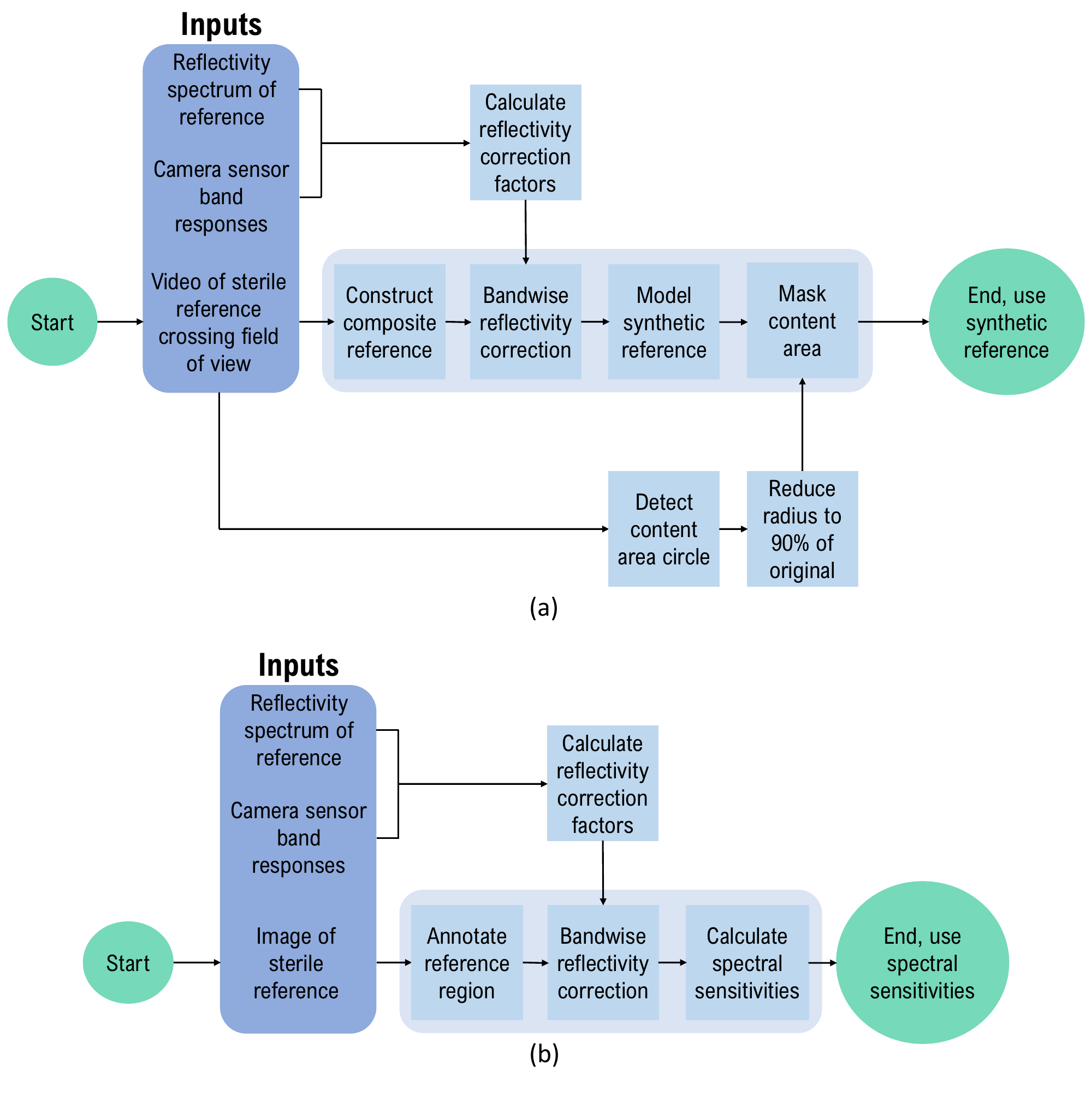}
 \caption{Flow charts demonstrating the algorithms for using a sterile reference to white balance intra-operative data for quantitative (a) or relative (b) spectral reconstructions.}
 \label{fig:algorithm}
\end{figure}

An overview of the resulting computational pipeline to convert a short video of the ruler to a synthetic white reference is shown \figref{fig:algorithm}a. This allows reconstruction of quantitatively accurate spectral data. In some cases only relative data is required where a simplified algorithm can be used to generate only spectral sensitivities, as shown in \figref{fig:algorithm}b. Individual steps are detailed in the remainder of the section.

\subsubsection{Generating composite image}
\label{algorithmcomposite}
Hyperspectral videos capture information across space, wavelength and time.
The video frames are stacked to create a spatio-spectral $x,y$ plane due to the mosaic structure, and a temporal $z$ axis.
Because of the sweeping motion of the bright ruler,
for each spatio-spectral pixel in the $x, y$ plane there is a temporal distribution of measured intensities, which alternates between background and sterile reference intensities in the $z$ axis.
The temporal regions corresponding to the sterile reference are isolated, as detailed below, and the median of these chosen as the value for that spatio-spectral location in the composite reference.
To segment these temporal regions,
we first pre-process the per pixel temporal intensity profile to suppress the impact of background values to a large extent by setting to 0 all the values whose intensity lie below a threshold computed with a classical parameter-free Otsu's method
\cite{Otsu1979} (using skimage v0.19.2 \href{https://scikit-image.org/docs/stable/api/skimage.filters.html\#skimage.filters.threshold_otsu}{\texttt{skimage.filters.threshold\_otsu}} function.)
We further clamp high intensities close to saturation with a fixed threshold in order to remove the impact of specular reflections.
These pre-processed temporal profiles are then smoothed using a Savitsky-Golay filter, with window size 15 and order 2
\cite{Savitzky1964a}
(using SciPy v1.8.1 \href{https://docs.scipy.org/doc/scipy/reference/generated/scipy.signal.savgol_filter.html}{\texttt{scipy.signal.savgol\_filter}} function),
to limit the impact of noise.
Temporal gradients are then calculated for the resulting pre-processed intensities.
Finally, the peaks, detected using a peak picking algorithm which finds all local maxima by simple comparison of neighboring values and refines these based on their height and prominence (using SciPy v1.8.1 \href{https://docs.scipy.org/doc/scipy/reference/generated/scipy.signal.find_peaks.html}{\texttt{scipy.signal.find\_peaks}} function),
are used to identify the temporal regions of interest.
The median of the raw values in these regions, prior to smoothing and thresholding, is then recorded and used as the estimated intensity of that spatial location in the initial mosaic composite reference $W^{c,m}(x, y)$.
%
%
%
%
%
%
%
%
%

\subsubsection{A priori ruler reflectivity correction}
\label{algorithmreflectivity}
\begin{figure}[h!]
	\centering
	\includegraphics[width=0.7\textwidth]{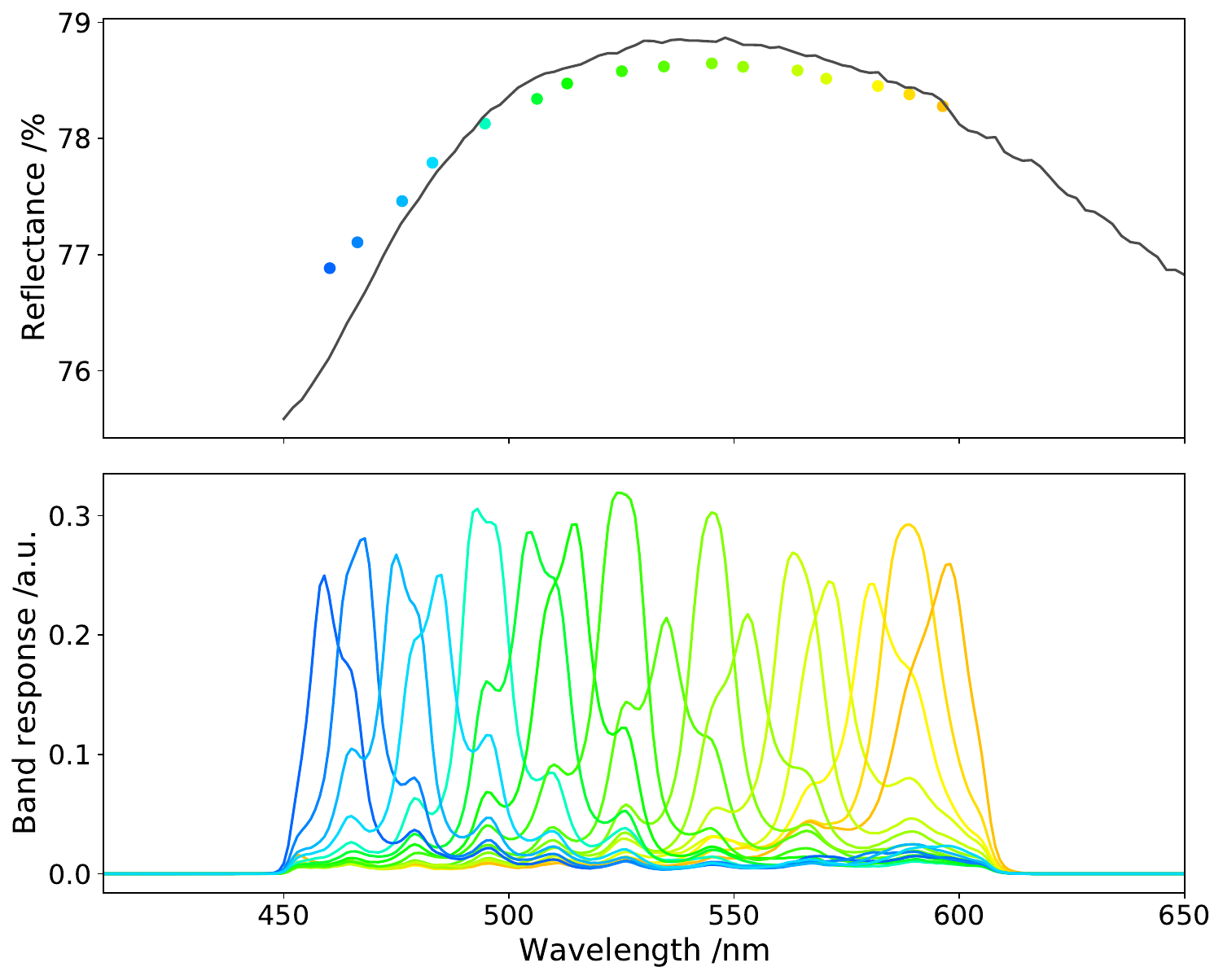}
	\caption{The reflectance spectrum of the ruler surface without markings and the appropriate reflectivity correction factors computed from this (top) and the band responses of the camera (bottom).}
	\label{fig:rulerspectrum}
\end{figure}

The sterile ruler was measured in the laboratory using a benchtop spectrophotometer to obtain its reflectance spectrum $t(\lambda)$ as seen in \figref{fig:rulerspectrum}.
This ruler back can be approximated as a Lambertian surface as variation in spectrum due to small angle changes is small, and the images of this ruler appear matte with few specular reflections. The ideal imaging condition is fronto-parallel imaging, so angle deviations from this are considered sufficiently small as to have negligible effect. 
From there,
an a priori reflectivity correction factor $\rho^{a}_{n}$ is calculated per band $n$ of the mosaic sensor:
\begin{linenomath*}
\begin{equation}
	\rho^{a}_{n} = \frac{\int t(\lambda)b_n(\lambda) d\lambda}{\int b_n(\lambda) d\lambda}
\label{eq:reflectivitycorrection}
\end{equation}
\end{linenomath*}
where $b_n(\lambda)$ is the band response as provided by the sensor manufacturer. Both the reflectivity correction factors and the band responses can be seen in \figref{fig:rulerspectrum}. %

Each pixel of the initial composite reference $W^{\textrm{ic}}(i,j,n)$ is then divided by the a priori correction factor $\rho^{a}_{n}$ for the appropriate band to form the final composite reference: $W^{\textrm{c}}(i,j,n) = W^{\textrm{ic}}(i,j,n)/\rho^{a}_{n}$.
We note that this correction captures the sensor band responses and the spectrum of the ruler as known a priori but does not account for the unknown spectrum of the light source used intraoperatively. This provides an essential step in the generation of synthetic references or the spectral sensitivities. 
In the generation of synthetic references, this correction is performed on a composite reference which may also present important imperfections due to noise and video input processing. We aim to mitigate these through our separable model fitting.
%
%

\subsubsection{Separable model parameter fitting}
\label{algorithmparameters}
A variety of methods can be used to calculate the factors in \eqref{eq:modelling white references}. We present three options which vary in their modelling of the vignetting, and whether the spectral sensitivities and scalar factor are fitted jointly or sequentially. For simplicity, we present the methods given a measured white reference $W(i,j,n)$ as input although they can also be applied to a composite white reference $W^{\textrm{c}}(i,j,n)$.

\paragraph{Vignetting modelled non-parametrically}
The first method only makes use of the separability assumption, models the vignetting as a non-parametric function, and calculates the spectral sensitivities and scalar factor. Since vignetting is modelled as independent of wavelength, it should be identical for all bands.
Making use of our convention that $\frac{1}{N} \sum_{n}S(n) = 1$, and assuming a zero mean noise pattern $N(i,j,n)$, we find that, up to residual noise, \eqref{eq:modelling white references} leads to:
\begin{linenomath*}
\begin{equation}
	M V(i,j) \approx \frac{1}{N}\sum_{n} W(i,j,n)
\label{eq:nonpara1}
\end{equation} 
\end{linenomath*}
Making use of our second convention that $\max_{i,j}V(i,j) = 1$, we find that, up to residual noise:
\begin{align}
	M &\approx \max_{i,j}\left( \frac{1}{N}\sum_{n} W(i,j,n) \right) \\
	V(i,j) &\approx \frac{ \frac{1}{N}\sum_{n} W(i,j,n) }{ \max_{i,j}\left( \frac{1}{N}\sum_{n} W(i,j,n) \right) }
\end{align} 
Finally, spectral sensitivities can be estimated by:
\begin{linenomath*}
\begin{equation}
	S(n) \approx \frac{\sum_{i,j}W(i,j,n)}{\frac{1}{N}\sum_{i,j,n'}W(i,j,n')}
\label{eq:nonpara2}
\end{equation}
\end{linenomath*}
This method accounts for imperfections in the system such as dust on lenses more appropriately, however it is also more likely to retain imperfections from the input reference, for example artefacts in the compositing process. It is this \eqref{eq:nonpara2} that is used to calculate the spectral sensitivities for balancing of relative data as in \figref{fig:algorithm}b.

\paragraph{Vignetting modelled with a Gaussian function}
Alternatively, a parametric model can be used to capture the vignetting.
As is standard in computer vision, we use a two-dimensional, normalized, isotropic Gaussian for this purpose:
\begin{linenomath*}
\begin{equation}
	V(i,j) = \exp\left(-\frac{(i-\mu_i)^2}{2\sigma^2} - \frac{(j-\mu_j)^2}{2\sigma^2} \right)
\label{eq:2DGauss}
\end{equation}
\end{linenomath*}
where $\mu_i$ and $\mu_j$ represent the co-ordinates of the centre of the Gaussian in the $i$ and $j$ directions respectively and $\sigma$ represents the standard deviation. 

Our second method fits the Gaussian parameters, the scalar factor and spectral sensitivities using a joint least squares approach:
\begin{linenomath*}
\begin{equation}
	\argmin_{\mu_i,\mu_j,\sigma, M, \{S(n)\}_{n}} \sum || W(i,j,n) - MS(n) \exp\big( -\frac{(i-\mu_i)^2}{2\sigma^2} - \frac{(j-\mu_j)^2}{2\sigma^2}\big) ||^2
\end{equation}
\end{linenomath*}

The third method calculates the scalar factor $M^{\textrm{est}}$ and spectral sensitivities $S^{\textrm{est}}(n)$ as in the first method with the Gaussian fitted to the non-parametric vignetting using a 
least squares approach on the Gaussian parameters only:
\begin{linenomath*}
\begin{equation}
	\argmin_{\mu_i,\mu_j,\sigma} \sum || W(i,j,n) - M^{\textrm{est}}S^{\textrm{est}}(n) \exp\big( -\frac{(i-\mu_i)^2}{2\sigma^2} - \frac{(j-\mu_j)^2}{2\sigma^2}\big) ||^2
\end{equation}
\end{linenomath*}
As will be apparent from our experiments, this third method provides advantageous benefits due to the Gaussian fit mitigating imperfections in the composite reference, whilst minimising the computational time.

\paragraph{Masking content area}
Finally, as detailed in \citenum{munzer13,huber22}, imaging through a scope leads to a circular content area which needs to be taken in to account in our computational pipeline.
In this work, the content area disk
is detected from a frame of a video using a pre-trained neural network~\cite{huber22} (using the implementation at \url{https://github.com/RViMLab/endoscopy}.)
The radius is reduced to 90\% of its original size to discount edge effects. This circle is then used to mask the content area as the final step to generate the synthetic white reference.

%
%
%
%
%
%

%
%
%
%
%
%
%

%
%
%
%
%
%
%
%
%
%
%
%

%
%
%
%

%

\subsubsection{Balancing using only spectral sensitivity for relative data}
\label{algorithmrelative}
For some use cases, absolute reflectance information $R(i,j,n)$ is not required so relative data is sufficient as described in \secref{intro}.
When normalising the spectral data at each pixel after \eqref{eq:WBfull} using \newline $R^{\textrm{norm}}(i,j,n) = R(i,j,n) / \frac{1}{N}\sum_{n'}|R(i,j,n')|$, the spatial effects of the vignetting $V(i,j)$ and absolute intensity information arising from $M$ is lost.
In this context, only spectral sensitivities are required for white balancing.
Dismissing the effect of dark images for simplicity, this is performed as follows:
\begin{linenomath*}
\begin{equation}
    R^{\textrm{norm}}(i,j,n) \approx \frac{I(i,j,n)}{S(n) \frac{1}{N}\sum_{n'}|\frac{I(i,j,n')}{S(n')}|} 
\end{equation}
\end{linenomath*}
However this removes the possibility of proper sRGB or CIELAB construction.
The simplified algorithm using only a single image of the ruler to generate these spectral sensitivities is also depicted in \figref{fig:algorithm}b. 

\subsection{Evaluation methodology for synthetic references}
\label{methodsynthetic}
For each distance in our experimental setup, a video of the ruler moving across the field of view was taken against a static background.
These were used to construct synthetic white references for each distance using the model described in \secref{methodmodel} with Gaussian vignetting and calculated scalar factor and spectral sensitivities.
Synthetic references were then used to reconstruct quantitative spectra from our checkerboard datasets and the errors calculated as in 
\secref{methodmotivation}.

The similarity of the synthetic references to measured references was evaluated using a range of metrics.
The pixel-by-pixel absolute percentage errors between the synthetic and measured references, and the absolute percentage errors between each set of spectral sensitivities were summarized in terms of median absolute percentage error ($\MedAPE$):
\begin{linenomath*}
\begin{equation}
    \MedAPE\big(u(\cdot),u^{\textrm{ref}}(\cdot)\big) = \median_{k}\Big|\frac{u(k) - u^{\textrm{ref}}(k)}{u^{\textrm{ref}}(k)}\Big|
\label{eq:abspercenterr}
\end{equation}
\end{linenomath*}
where $u$ is a given signal compared against a given reference $u^{\textrm{ref}}$ both indexed by $k$.

For the evaluation of the relative spectra achieved with the method presented in \secref{algorithmrelative},
a single set of spectral sensitivities is calculated based on the lighting conditions from a 150 pixel radius region of interest in a single image of a ruler.
These spectral sensitivities are then used to reconstruct the relative spectra at all distances and the errors calculated in these as in \secref{methodmotivation}.
However, good sRGB and CIELAB reconstructions cannot be computed from relative data as the intensity information is lost so this analysis is not performed in this case.

Finally, the synthetic reference method is evaluated in terms of surgical workflow integration in a sterile environment by its use in a human cadaveric spine surgery. Qualitative evaluation is performed on sRGB reconstructions as part of \secref{resultsintegration}.
%

\section{Results}
\label{results}
\subsection{Impact of improper white balancing}%
\begin{figure}[hp!]
	\centering
        \includegraphics[width=0.8\textwidth]{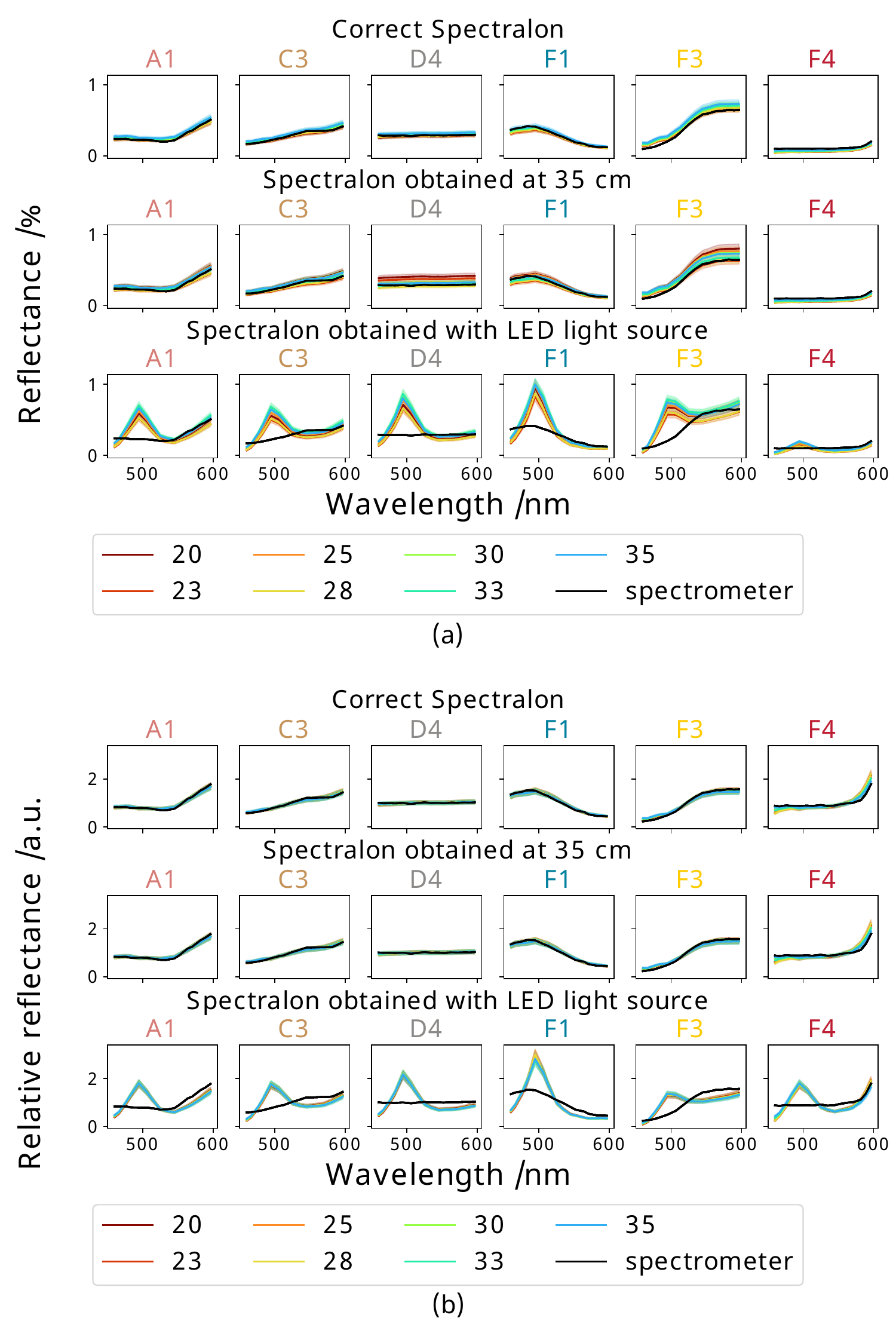}
 \caption{The mean spectrum of a selection of tiles as measured from each distance (cm) plotted against the spectrometer measurement of that tile, where each plot represents a new tile (labelled above it as corresponding to those in \figref{fig:setup}b) for both (a) quantitative and (b) relative data. At each distance the data is balanced with either the correct Spectralon white reference image, the Spectralon reference obtained at 35cm, or the Spectralon reference obtained using a different light source.}
 \label{fig:summaryspectranecessity}
\end{figure}

\begin{figure}[t!]
	\centering
        \begin{subfigure}[htp!]{0.55\textwidth}
	    \includegraphics[width=\textwidth]{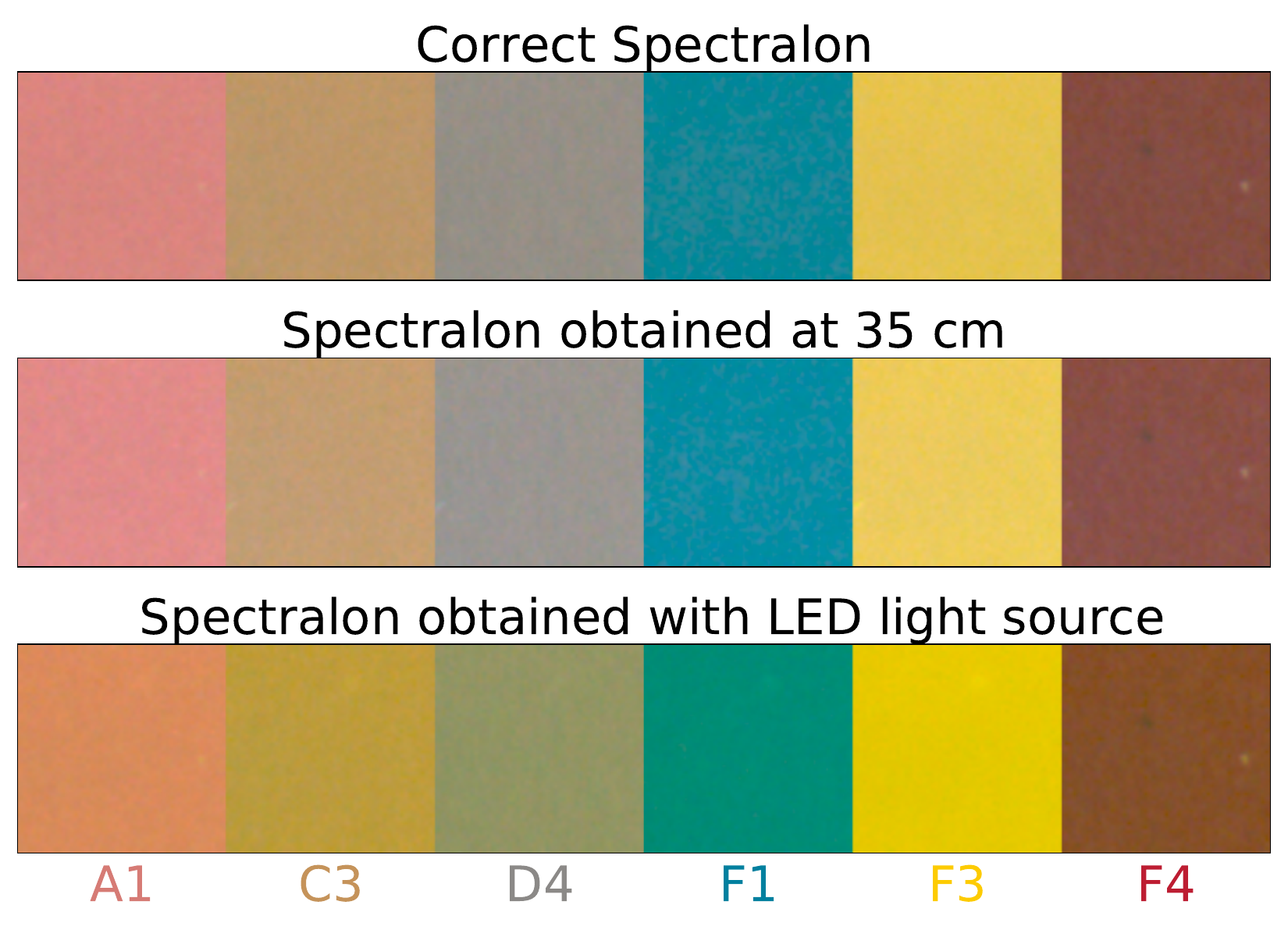}
	    \label{fig:20Necessity}
	\end{subfigure}
 \caption{sRGB reconstructions of the central 200 x 200 pixels of a selection of tiles (labelled below as corresponding to those in \figref{fig:setup}b) taken at 20 cm and balanced with either the correct Spectralon white reference image, the Spectralon reference obtained at 35cm, or the Spectralon reference obtained using a different light source.}
	\label{fig:summarysRGBnecessity}
\end{figure}

\label{resultsnecessity}
Datasets were obtained at a variety of distances and under different lighting conditions as described in \secref{methodmotivation}. 
To summarize this spectral data, six representative tiles of interest (A1, C3, D4, F1, F3, F4) are displayed for the variety of white balancing regimes investigated in \figref{fig:summaryspectranecessity}, and the sRGB reconstructions of each regime for the dataset at 20 cm shown in \figref{fig:summarysRGBnecessity}. These tiles demonstrate primary colors and some colors expected in surgical scenes. The associated errors for these data are shown in \tabref{tb:badwhites}. 

The first regime shows the optimally reconstructed spectra for a variety of distances of the camera from the checkerboard. These are calculated by balancing each dataset obtained using the Karl Storz D light C using a measured Spectralon white reference obtained at the same location as the data with the same light source, which is not possible in a surgical environment.

To demonstrate the impact of the white reference being obtained at a different distance to the subject, the checkerboard data at all distances is balanced with the white reference obtained at 35 cm. Visually a wider spread of intensities in quantitative spectra can be seen, and the errors confirm that as the distance varies from that of the white reference the errors in the quantitative reconstructed spectra also increase, as seen in \figref{fig:summaryspectranecessity}. The $\Delta E$ values and sRGB reconstructions do not show a significant difference compared to the correctly balanced spectra as a large amount of spectral information is lost on conversion to sRGB or CIELAB which reflects the increase in information expected from the use of hyperspectral imaging compared to conventional cameras. 

Another likely problem with limited white referencing in a surgical setting is the likelihood of changing lighting conditions. This was mimicked by obtaining a white reference with the Karl Storz LED light source at each distance and using this to balance the data collected with a different light source, the Karl Storz D light C, which was then processed as above. The resulting errors, spectra, and sRGB reconstructions clearly show the importance of white referencing to account for the lighting conditions of the scene as the shape of the spectra are significantly changed in both quantitative and relative spectra which is also clearly visible in the sRGB reconstruction and all error metrics so any change in lighting must be accounted for. 
\begin{table}[tbph]
    \caption{The mean $\NRMSE$ for the mean measured spectrum of each tile compared to its spectrometer spectrum of each quantitative and relative dataset, alongside the median pixel-by-pixel $\Delta E$ value between the composite CIELAB reconstructions compared to the literature values. These datasets are balanced with the correct Spectralon reference, the Spectralon reference obtained at 35cm, or the Spectralon reference obtained using an LED light source.}
    \csvloop{
file = BadWhites.csv, 
head to column names,
before reading = \centering\sisetup{round-mode=figures, round-precision=3},
tabular={|l|SSS|SSS|SSS|@{}c},
table head = \hline 
& \multicolumn{3}{c}{\textbf{$\NRMSE$ for Quantitative}} & \multicolumn{3}{|c|}{\textbf{$\Delta E$}} & \multicolumn{3}{c|}{\textbf{$\NRMSE$ for Relative}} \\
 & \multicolumn{3}{c}{\textbf{spectra (\%)}} & \multicolumn{3}{|c|}{ } & \multicolumn{3}{c|}{\textbf{spectra (a.u.)}} \\
\rot{\textbf{Distance (cm)}} & \rot{\textbf{Correct Spectralon}} & \rot{\textbf{Spectralon at 35 cm}} & \rot{\textbf{Spectralon using LED}} & \rot{\textbf{Correct Spectralon}} & \rot{\textbf{Spectralon at 35 cm}} & \rot{\textbf{Spectralon using LED}} & \rot{\textbf{Correct Spectralon}} & \rot{\textbf{Spectralon at 35 cm}} & \rot{\textbf{Spectralon using LED}} \\ 
\hline, 
command = \Distance & \qNRMSECorrect & \qNRMSEdist & \qNRMSELED & \qDeltaECorrect & \qDeltaEdist & \qDeltaELED & \nNRMSECorrect & \nNRMSEdist & \nNRMSELED , 
table foot = \hline,
    }
\label{tb:badwhites}
\end{table}

%
%
%
%
%
%
%
%
%
%
%
%
%
%
%
%
%
%
%
%
%
%
%
%
%
%
%
%
%
%
%
%
%
%
%
%
%
%
%
%
%
%
%
%
%
%
%
%
%
%
%
%
%
%
%
%
%
%
%
%
%
%
%
%
%
%
%
%
%
%
%
%
%
%
%
%
%
%
%
%
%
%
%
%
	%
	%
	%
	%
	%
    %
%
%
%
%
%
%
%
%
%
	%
	%
	%
	%
	%
	%
	%
	%
	%
	%
	%
	%
	%
	%
 %
    %
	%
 %
%
%
%

\FloatBarrier
\subsection{Goodness of fit of the separable white reference models}
\label{resultsmodel}
\begin{figure}[h!]
	\centering
	\includegraphics[width=0.95\textwidth]{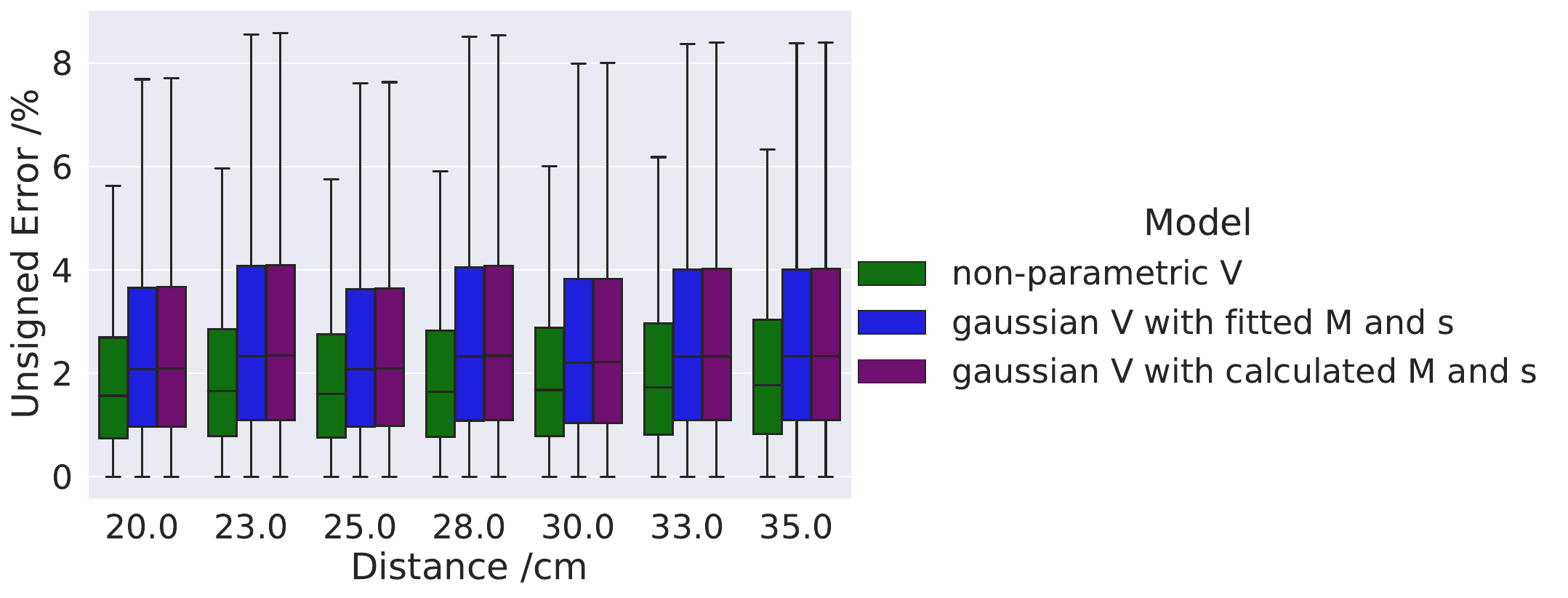}
	\caption{Box plots showing the pixel-by-pixel absolute percentage errors in modelled references compared to the original references at a range of distances using either non-parametric, Gaussian vignetting with fitted scalar factor and spectral sensitivities, and Gaussian vignetting with calculated scalar factor and spectral sensitivities methods.}
	\label{fig:ModelErrors}
\end{figure}

The measured white references obtained at each distance were approximated using the non-parametric, Gaussian vignetting with jointly-fitted scalar factor and spectral sensitivities, and Gaussian vignetting with calculated scalar factor and spectral sensitivities methods to determine the inherent errors in using this model and these approximations.
The pixel-by-pixel absolute percentage errors in the resulting approximated references compared to their respective measured references are then plotted in \figref{fig:ModelErrors} and show very low errors across all models.
The models using Gaussian fitting show slightly higher errors, likely due to the inability of the Gaussian fit to account for dust on lenses in the set-up. 

The spectral sensitivities obtained for each model were found to be invariant with respect to distance or approximation method. Any change in distance requires a change in optical focus which results in a change of optical vignetting. This suggests that the spectral sensitivities are unaffected by optical vignetting or its modelling. This lends credence to the assumption that the spatial and spectral dependencies of the components of white references can be separated.

This confirms that white reference images can be well modelled using the assumption of separability between spatial and spectral components, and using a two-dimensional isotropic Gaussian.

\FloatBarrier
\subsection{Evaluation of synthetic references from sterile ruler sweeps}
\label{resultssynthetic}
\begin{figure}[htp!]
	\centering
         \begin{subfigure}[ht!]{0.6\textwidth}
	    \includegraphics[width=\textwidth]{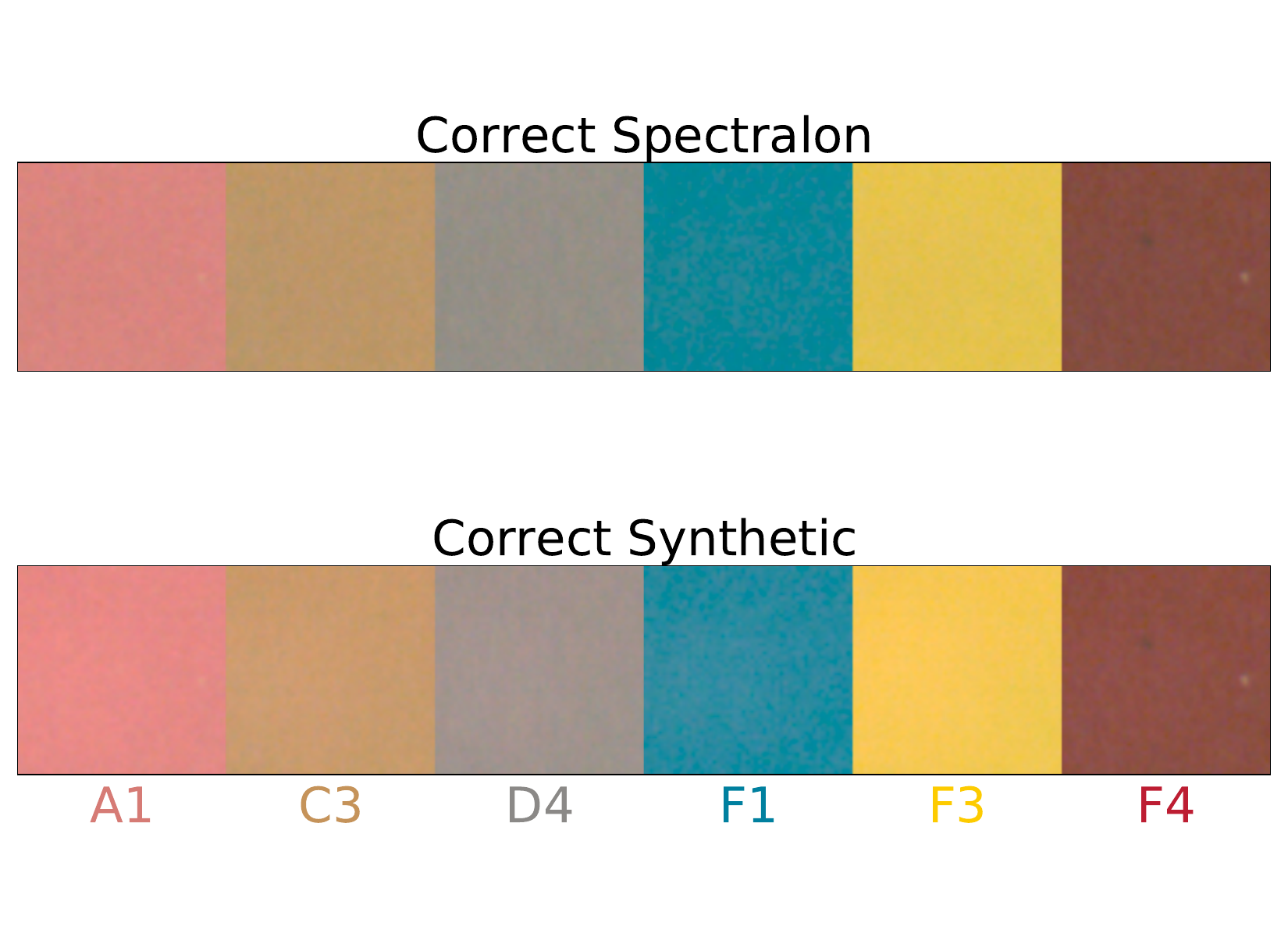}
	    \label{fig:20Synthetic}
	\end{subfigure}
 \caption{sRGB reconstructions of the central 200 x 200 pixels of a selection of tiles (labelled below as corresponding to those in \figref{fig:setup}b) taken at 20 cm and balanced with either the correct Spectralon white reference image, or the synthetic reference generated from the correct video of the sterile ruler.}
	\label{fig:summaryspectrasynthetic}
\end{figure}

The errors between the synthetic references compared to measured references were calculated as described in \secref{methodsynthetic} and can be seen in \tabref{tb:SynTotErrors} for the synthetic references generated from videos of the ruler against a fixed background obtained at the same position as the data. These show that this method reproduces the measured Spectralon references well, particularly as the errors are comparable to the inherent errors in modelling references shown in \figref{fig:summaryspectrasynthetic}.
\begin{figure}[htp!]
	\centering
        \includegraphics[width=0.8\textwidth]{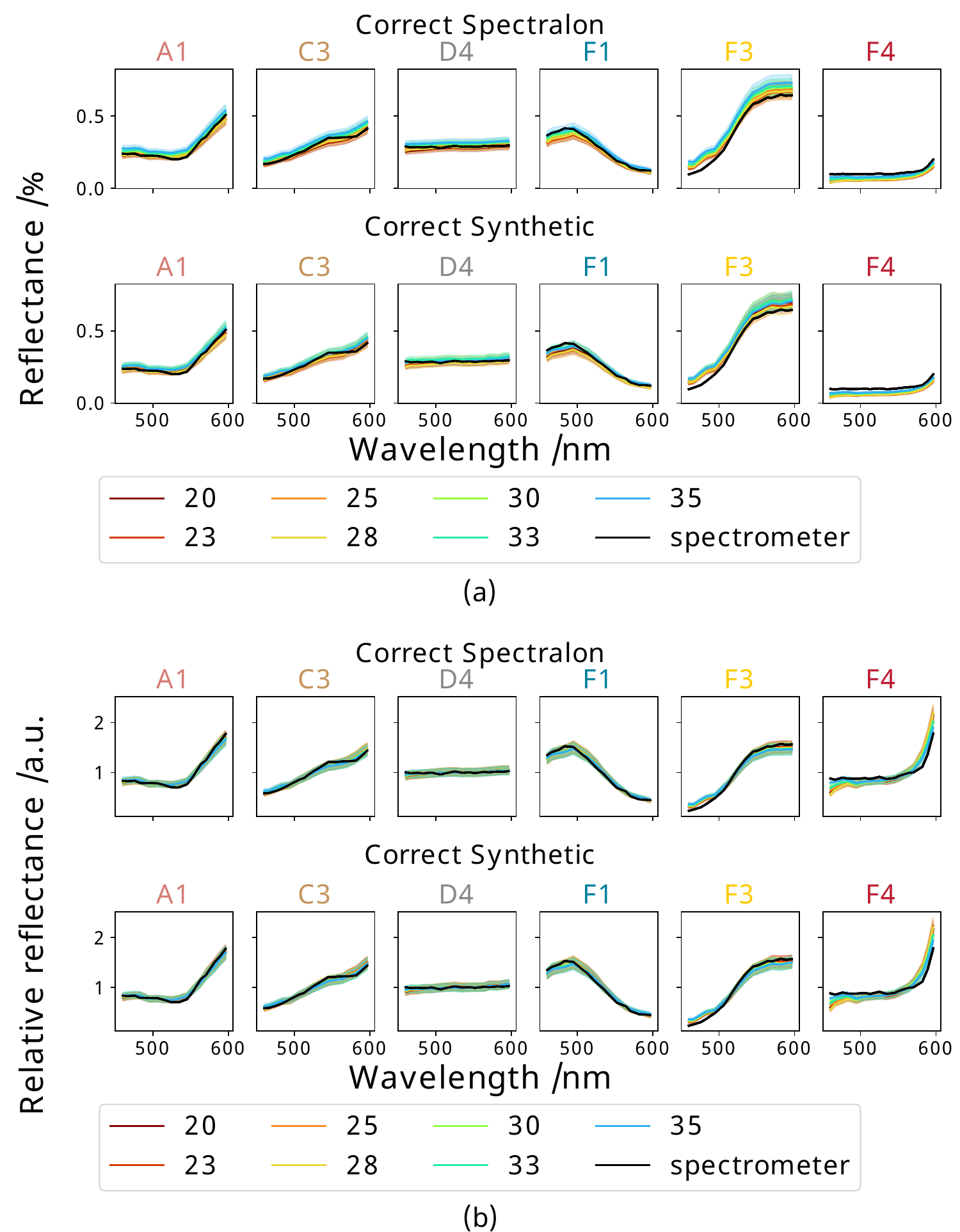}
 \caption{The mean spectrum of a selection of tiles as measured from each distance (cm) plotted against the spectrometer measurement of that tile, where each plot represents a new tile (labelled above it as corresponding to those in \figref{fig:setup}b) for both (a) quantitative and (b) relative data. At each distance the data is balanced with either the correct Spectralon white reference image, or the synthetic reference generated from the correct video of the sterile ruler.}
\end{figure}

\begin{table}[tbh!]
	\caption{Table showing the MedAPE (3sf) in the spectral sensitivities and in the pixel-by-pixel percentage errors (3sf) between the synthetic reference generated from a video of the sterile ruler and the reflectivity corrected Spectralon reference for each distance, alongside the mean $\NRMSE$ (3sf) when balancing with each synthetic reference for the quantitative and relative datasets for the mean measured spectrum of each tile compared to its spectrometer measurement, alongside the median pixel-by-pixel $\Delta E$ value (3sf) between the composite checkerboard CIELAB reconstruction and the literature values.}
	\csvloop{
		file = CombinedSyn.csv, 
		head to column names, 
		before reading = \centering\sisetup{table-alignment-mode=none, table-number-alignment=center, round-mode=figures, round-precision=3, separate-uncertainty=true},
		tabular={|l|S|S|S|S|S@{}c|}, 
		table head = \hline
		\textbf{Distance} & \multicolumn{2}{|c}{\textbf{MedAPE (\%)}} & \multicolumn{2}{|c|}{\textbf{$\NRMSE$}} & \textbf{$\Delta E$} & \\
		 \textbf{(cm)} & \textbf{$S_n$} & \textbf{pixel-by-pixel} & \textbf{Quantitative (\%)} & \textbf{Relative (a.u.)} &  & \\
		\hline, 
		command = \Distance & \Checkerss & \Checkerpbp & \quantNormRMSEa &  \normNormRMSEa & \quantmedianDeltaE & , 
		table foot = \hline , 
	}
	\label{tb:SynTotErrors}
\end{table}
\newpage
These synthetic references are also used to balance the checkerboard data as in \secref{methodmotivation} and the corresponding errors calculated. The results are summarized in \figref{fig:summaryspectrasynthetic}, and \tabref{tb:SynTotErrors} shows the resulting errors when using synthetic references. These show very similar errors compared to the best practice method suggesting that these synthetic references are suitable for quantitative data reconstruction. Since this process is fully sterile, it can be repeated easily during surgery allowing for changes in distance and lighting conditions to be better accounted for.

\begin{figure}[htp!]
	\centering
        \includegraphics[width=0.8\textwidth]{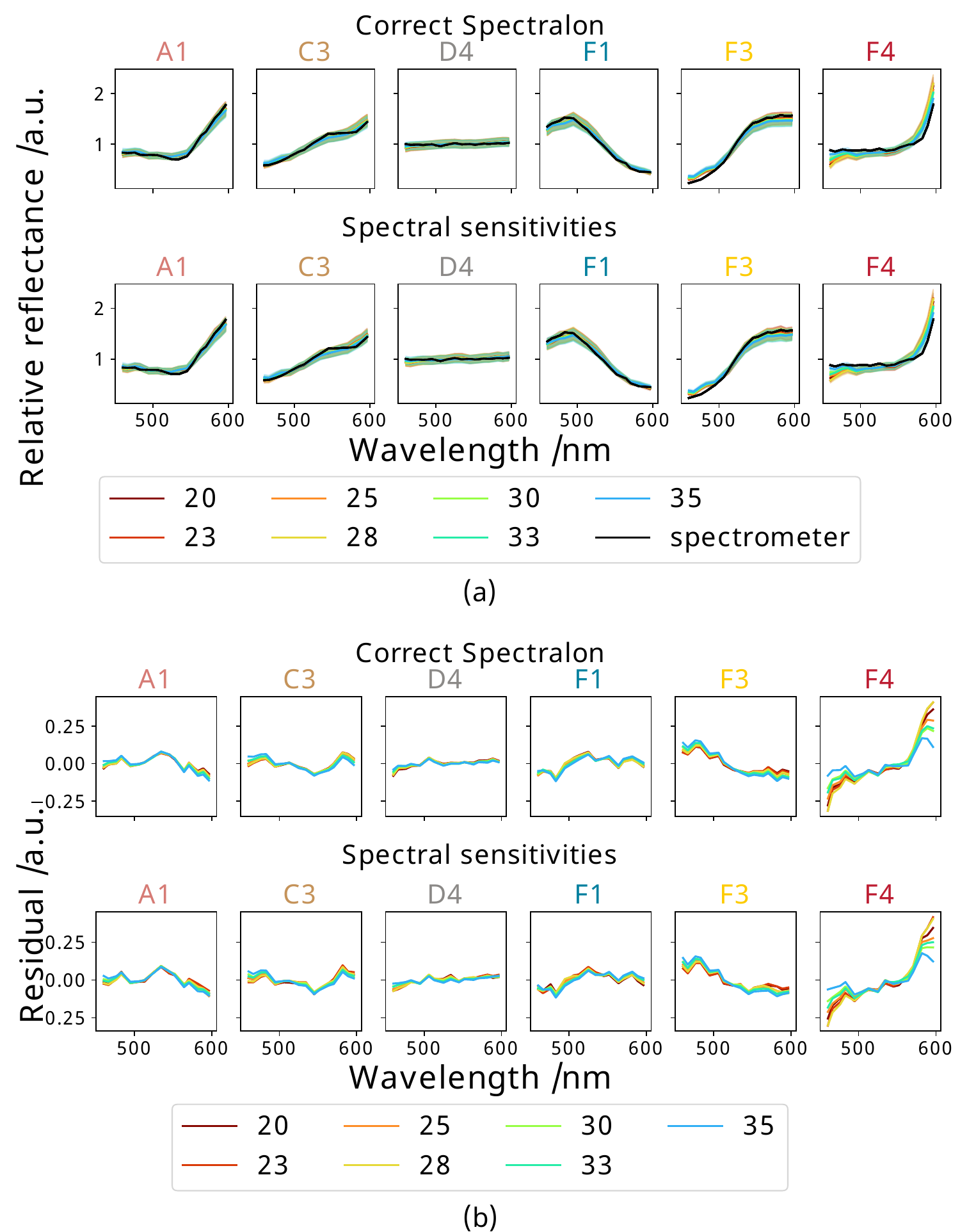}
	\caption{(a) The relative mean spectrum of a selection of tiles as measured from each distance (cm) plotted against the spectrometer measurement of that tile, where each plot represents a new tile (labelled above it) as corresponding to the labels in \figref{fig:setup}b, and (b) their respective residuals compared to the spectrometer ground truths. All distances of checkerboard data are white balanced with either the correct measured white reference or the spectral sensitivities generated from a single image of the sterile ruler.}
	\label{fig:NormSpecsens}
\end{figure}

\FloatBarrier
For relative data, the spectral sensitivities are calculated from a single image of the ruler and used to balance the data. The spectral sensitivities of the Karl Storz D light C Xenon light source are used to balance the data before normalising and the resulting spectra are summarized in \figref{fig:NormSpecsens} and the corresponding errors in \tabref{tb:NormSpecsens}. This shows that spectral sensitivities alone are sufficient for white balancing of relative data.
\begin{table}[htp!]
	\caption{Table showing the $\NRMSE$ (3sf) when balancing with the spectral sensitivities generated from a single image of a sterile ruler for the mean measured spectrum of each tile compared to its spectrometer measurement.}
	\csvloop{
		file = NormSpecsens.csv, 
		head to column names, 
		before reading = \centering\sisetup{table-alignment-mode=none, table-number-alignment=center, round-mode=figures, round-precision=3, separate-uncertainty=true},
		tabular={|l|S@{}c|}, 
		table head = \hline \textbf{Distance (cm)} & \textbf{$\NRMSE$ for} & \\
	& \textbf{Relative Spectra (a.u.)} & \\
	\hline, 
		command = \Distance & \normNormRMSEa & , 
		table foot = \hline, 
	}
	\label{tb:NormSpecsens}
\end{table}

\FloatBarrier
\subsubsection{Integration into a surgical environment}
\label{resultsintegration}
\begin{figure}[h!]
	\centering
	\includegraphics[width=0.56\textwidth]{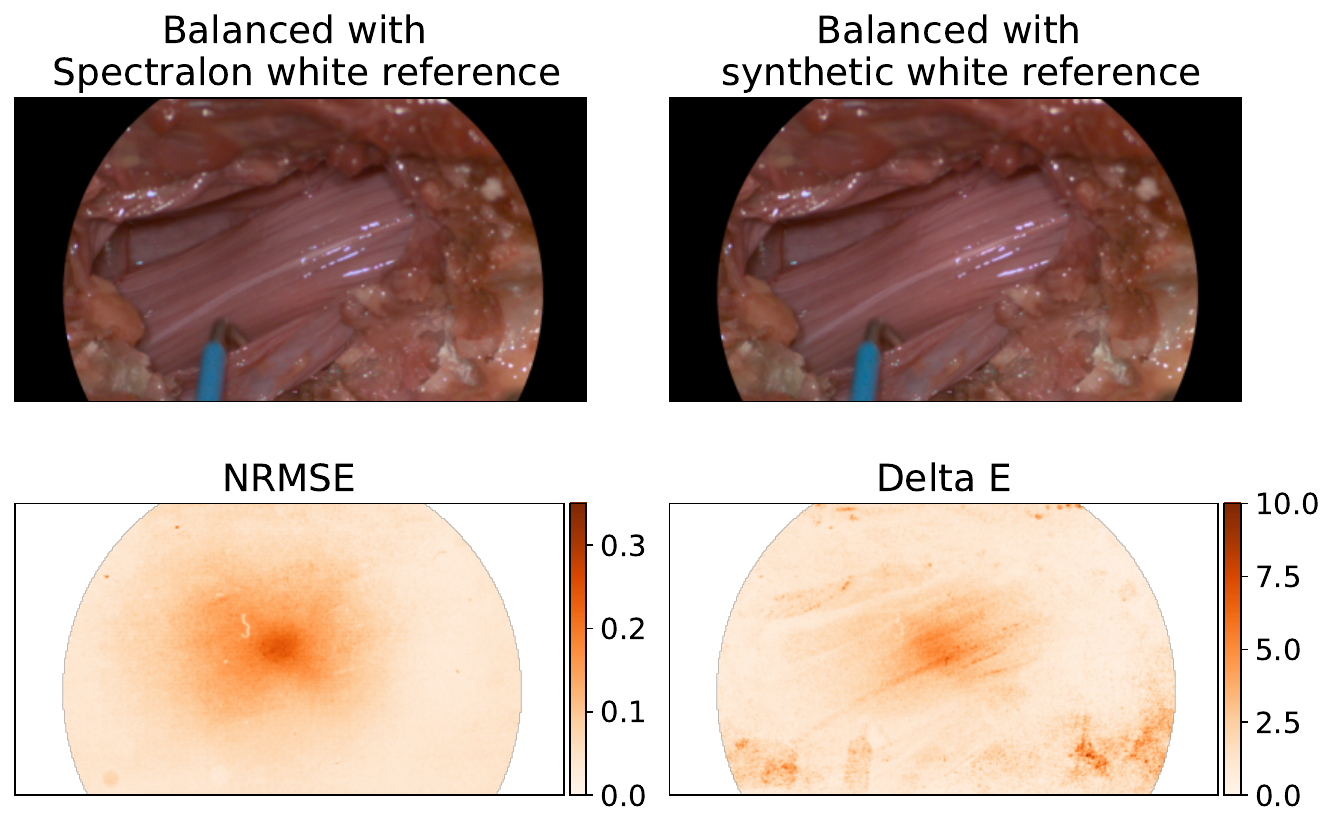}
	\caption{sRGB reconstructions of human cadaveric rootlets balanced with a Spectralon reference and a synthetic white reference generated from a video of a sterile ruler against tissue, the pixel-by-pixel $\NRMSE$, and $\Delta E$ values calculated between the balanced spectra and CIELAB reconstructions of these. All images are masked due to the scope.}
	\label{fig:ZurichRGB}
\end{figure}

Our proposed synthetic reference algorithm was easily integrated into surgical workflow of a human cadaveric spine surgery. The study was approved by the local ethics committee of the canton of Zurich, Switzerland, under the number BASEC 202-01196. The video capture was adjusted to use a snipped section of the ruler to fit within the surgical cavity and the camera was moved in contrast to the ruler to capture the video. In this setting this is an appropriate adjustment due to the near uniformity of the background tissue and even illumination of the cavity using a single light source.
The measured white reference was taken for comparison, where distance was judged based on a visual assessment of image focus. The mean absolute percentage errors in the spectral sensitivities is 0.852\% and median absolute percentage pixel-by-pixel errors is 4.77\% between the synthetic and Spectralon references, which show that the synthetic reference is very similar to the best case intraoperative measured Spectralon reference. The $\NRMSE$ between the quantitative spectra constructed using the Spectralon and synthetic references were calculated between the spectrum of each pixel in the content area. This had a range of 0\% to 0.356\% with a median of 0.055\% showing that these are extremely similar. sRGB reconstructions of the image balanced with both Spectralon white reference and a synthetic white reference are shown in \figref{fig:ZurichRGB}. The $\Delta E$ values were calculated for each pixel between CIELAB reconstructions of these images and displayed in \figref{fig:ZurichRGB} with a range between 0 and 16.17 and a median of 1.37 in the content area indicating a barely perceptible difference to the human eye. This suggests that a synthetic white reference performs similarly to the best case intraoperative Spectralon white reference, however it is much more easily integrated into surgical workflow allowing for changing conditions to be better accounted for. 

\FloatBarrier
\section{Discussion}
\label{conclusions}
This study first examined the limitations of white balancing under different conditions to the subject. The lack of a sterile reference necessitates that the white reference be taken outside of the sterile field. This means that the white reference may be taken at a different distance or under different lighting conditions to the subject. When investigating these effects, it is clear that there is little variation when data are correctly balanced, however changes in distance or lighting from the conditions of the white reference show appreciable differences. 

The change in distance primarily has an effect on quantitative data which is seen clearly at large differences in distance, however these changes are not seen in the sRGB reconstructions. The relative data is robust to changes in distance as this does not change the shape of the spectra, only the absolute intensity which is not relevant in calculation of ratiometric parameters. 

The change in lighting conditions has a severe effect on both quantitative and relative data as the shape of the spectra are altered. This effect is significant enough to be easily visually apparent in the sRGB reconstruction. 

The ability to model white references by separating the spatial and spectral components was then investigated. This produced relatively low inherent errors for all methods investigated. The approximation method modelling the vignetting as a two-dimensional isotropic Gaussian with the scalar factor and spectral sensitivities calculated was chosen to allow for imperfections in the composite reference whilst maximising computational efficiency. The spectral sensitivities in all three approximation methods showed no trends with distance, consistent with the assumption that the spatial and spectral effects can be treated separately. 

Synthetic references were generated from short videos of the ruler and compared to the relevant measured references. The errors in the spectral sensitivities and pixel-by-pixel values of the synthetic references compared to the appropriate measured references, of less than 6.5\%, suggest that there is little difference between the measured and synthetic references. 

When these synthetic references were used to balance the datasets at various distances, the results appeared similar to using appropriate measured references, with mean $\NRMSE$ for the quantitative spectra of 0.140\% and 0.139\% respectively. This demonstrates that these synthetic references can be used as an appropriate method of white balancing to obtain quantitatively accurate data. 

For relative data, only the shape of the spectra is relevant so only spectral sensitivities are required to white balance the data. For this the relevant spectral sensitivities are calculated once and used to balance the data across all datasets which results in very low errors. This shows that this simplified method of white balancing can be used in the case that relative data alone is required, however it is not possible to generate good sRGB reconstructions with only spectral sensitivities, since absolute intensity information is lost. 

The application of this method in a surgical environment was tested in a human cadaveric spine surgery. This demonstrated a simple integration into a surgical environment whilst producing similar spectra and sRGB reconstructions, with median $\NRMSE$ of 0.055\% and median $\Delta E$ of 1.37, and a synthetic reference that closely replicates the best case scenario using a standard reference. 

It is expected that this method can be integrated to a wide range of surgical applications where the exoscope can be held at a constant distance for the short ruler video, and the imaging of interest. This has been investigated for fronto-parallel distances of 20-35cm, which facilitates good ergonomics for neurosurgery by enabling the capture of a sufficiently wide field of view as well as resolving regions of interest in the depths of cavities, based on experience from the lead clinician on our neurosurgical study. Exposure should be set to prevent saturation of any wavelengths, however changes in exposure could be corrected linearly. If lighting is uniform across the field of view, this allows the video to be taken by movement of camera instead of reference, however shadows in the cavity can only be accounted for by movement of the reference itself.

This indicates a promising novel method for synthetic white balancing within the restricted intraoperative environment provided the assumption that spatial and spectral independence is held. This assumption is often met using a single, uniform light source. The quantitative model is currently limited to fronto-parallel imaging with the assumption that imaging geometry does not deviate significantly from this, which also introduces a limitation in its application, however relative data can be obtained at any angle in principle using the spectral sensitivities independently.

\subsection{Conclusions and Future Work}
These data demonstrate a novel sterile synthetic white balancing algorithm capable of integrating easily into surgical workflow. This ensures quantitative accuracy of the spectral data obtained, whilst minimising disruption to the surgical procedure.

Future work should include improvement of the computational efficiency of the composite ruler reference construction for a real-time integration of this method. It should be confirmed that this construction is robust to movement in the background as this enables the video to be obtained by movement of the camera in contrast to movement of the ruler, allowing this method to be more easily used in more restricted surgical cavities assuming an even illumination from a single light source. This method should also be adapted to be able to account for any dust in the system, multiple light sources, validation for the replacement of the exoscope with other operating scopes, validation for distances applicable to other open surgery geometries, and adaptation to use at different imaging angles.
\section*{Acknowledgements}
This study/project is funded by the NIHR [NIHR202114]. The views expressed are those of the author(s) and not necessarily those of the NIHR or the Department of Health and Social Care.
This work was supported by core funding from the Wellcome/EPSRC [WT203148/Z/16/Z; NS/A000049/1].
This project has received funding from the European Union's Horizon 2020 research and innovation programme under grant agreement No 101016985 (FAROS project).
CH is supported by an InnovateUK Secondment Scholars Grant (Project Number 75124).
TV is supported by a Medtronic / RAEng Research Chair [RCSRF1819\textbackslash7\textbackslash34].
For the purpose of open access, the authors have applied a CC BY public copyright licence to any Author Accepted Manuscript version arising from this submission.

\section*{Data availability statement}
The datasets presented in this article are not readily available because of their proprietary nature.

\section*{Disclosure statement}
ME, JS and TV are co-founders and shareholders of Hypervision Surgical.
%
\FloatBarrier
\typeout{}
\bibliography{Main_body}
\doublespacing
\listoffigures

%
\end{document}